\documentclass[reprint, longbibliography, superscriptaddress]{revtex4-1}
\usepackage{graphicx, graphics}
\usepackage{array}
\usepackage{enumitem}
\usepackage{wrapfig}
\usepackage{hyperref}
\hypersetup{colorlinks,allcolors=blue}
\usepackage{amsmath, amssymb}
\usepackage{mathtools}
\usepackage{listings}
\usepackage{xcolor} 
\usepackage{colortbl}
\usepackage{mdframed}
\definecolor{Gray}{gray}{0.85}

\usepackage{makecell, cellspace, caption}
\usepackage{array}
\newcolumntype{L}[1]{>{\raggedright\let\newline\\\arraybackslash\hspace{0pt}}m{#1}}
\newcolumntype{C}[1]{>{\centering\let\newline\\\arraybackslash\hspace{0pt}}m{#1}}
\newcolumntype{R}[1]{>{\raggedleft\let\newline\\\arraybackslash\hspace{0pt}}m{#1}}

\definecolor{mygray}{gray}{0.9}
\newcolumntype{g}{>{\columncolor{mygray}}c}

\usepackage{multirow, makecell}
\usepackage{xcolor}
\usepackage{booktabs}
\usepackage{ragged2e}
\usepackage{tabularx}
\usepackage{enumitem}
\usepackage{csquotes}
\usepackage{pythonhighlight} 

\usepackage{color} 
\usepackage{framed}

\definecolor{shadecolor}{gray}{0.95}

\newenvironment{shadedquotebox}{
    \begin{mdframed}[
        backgroundcolor=gray!10,
        linecolor=black,
        innerleftmargin=10pt,
        innerrightmargin=10pt,
        innertopmargin=10pt,
        innerbottommargin=10pt,
        leftmargin=0pt,
        rightmargin=0pt,
        align=center,
        userdefinedwidth=0.95\columnwidth
    ]
    \sffamily\slshape
}{
    \end{mdframed}
}

\makeatletter
\renewcommand\frontmatter@abstractwidth{\dimexpr\textwidth-1in\relax}
\makeatother
\usepackage{smartdiagram}

\begin{document}

\author{Andrew Low}
  \affiliation{Department of Physics, University of Livepool, Liverpool, UK, L69 7ZE} 
  \email{andrew.low@liverpool.ac.uk}%

\author{Z. Yasemin Kalender}
  \affiliation{Department of Physics, University of Livepool, Liverpool, UK, L69 7ZE} 
  \email{y.kalender@liverpool.ac.uk}%

\title{Data Dialogue with ChatGPT: Using Code Interpreter to Simulate and Analyse Experimental Data}

\begin{abstract}
Artificial Intelligence (AI) has the potential to fundamentally change the educational landscape. Chatbot's such as ChatGPT allow users to engage in conversations that mimic human interactions. So far, much of the physics education research relating to AI has focused on lecture-based assessment and the ability of ChatGPT to answer conceptual surveys and traditional exam-style questions. In this study, we shift the focus by investigating ChatGPT's ability to complete an introductory mechanics laboratory activity by using Code Interpreter, a recent plugin that allows users to generate and analyse data by writing and running Python code `behind the scenes'. By uploading a common `spring constant' lab activity using Code Interpreter, we investigate the ability of ChatGPT to interpret the activity, generate realistic model data, produce a line-fit, and calculate the reduced chi square statistic. By analysing our interactions with ChatGPT, along with the Python code generated by Code Interpreter, we assess how the quality and accuracy of ChatGPT's responses depends on different levels of prompt detail. We find that although ChatGPT is capable of completing the lab activity and generating plausible-looking data, the quality of the output is highly dependent on the detail and specificity of the text prompts provided. We find that the data generation process adopted by ChatGPT in this study leads to heteroscedasticity in the simulated data, which may be difficult for novice learners to spot. We also find that when \textit{real} experimental data is uploaded via Code Interpreter, ChatGPT is capable of correctly plotting and fitting the data, calculating the spring constant and associated uncertainty, and calculating the reduced chi square statistic. This work offers new insights into the capabilities of Code Interpreter within a laboratory setting and highlights a variety of text-prompt strategies for the effective use of Code Interpreter in a lab context.
 
\end{abstract}

\maketitle

\section{Introduction}\label{sec:intro}

In the last few years, artificial intelligence (AI) has emerged as a potentially revolutionary technology with multiple uses including natural language interpretation, machine learning, and complex data analysis. Perhaps the most famous example is ChatGPT, a large language model chatbot developed by OpenAI \cite{OpenAi} which, by January 2023, had become the fastest growing consumer software application in history, with over 100 million users \cite{SimilarWebBlog2023} (almost 60\% aged 18-34 \cite{SimilarWebDemographics}). The role that AI, especially software applications like ChatGPT, plays in higher education has been a topic of extensive research, with numerous studies exploring the implications for student learning, academic integrity, and traditional models of assessment \cite{Cotton2023_cheating, west2023ai, west2023advances, Kortemeyer_2023, Vasconcelos2023, article1234, Dahlkemper2023, Socrates2023, AISchool, DeathofPhysicsessay_Yeadon2023, kumar2023chatgpt4, Kung_medical2023, APcheating, ChatGPTinPER_Bitzenbauer}.

Over the past year, a wide range of ChatGPT plug-ins have been developed to overcome some of the fundamental constraints of Large Language Models (LLMs), and enhance the functionality and reliability of ChatGPT \cite{ChatGPTPlugins}. In July 2023, ChatGPT released a beta version of Code Interpreter, which enables users to upload and download files, analyse data, perform mathematical calculations, and create and interpret code, all of which are essential tools in STEM-related disciplines. Code Interpreter's ability to write and execute Python code `behind the scenes' and display the output within the standard ChatGPT user interface has been hailed as a game-changer for data analysis, and expands the range of possible applications of ChatGPT within an educational setting \cite{Lo2023EducationReview}.

In this study, we test some of the key features of the Code Interpreter plug-in. As a novelty in its approach and context, our study investigates the capability of Code Interpreter to interpret and complete an introductory physics laboratory activity within an undergraduate degree programme. In addition, this study aims to explore the broader implications for how Code Interpreter might impact teaching and learning within a laboratory context. Below, we will describe related literature on AI use in higher education, laboratory teaching, and our research questions for this study.

\section{Literature Review}\label{sec:literature}

\subsection{ChatGPT models}

 Large Language Models such as ChatGPT are capable of mimicking human conversation by generating responses to text prompts, after being trained on a vast quantity of textual material. ChatGPT is perhaps the most well-known LLM, and is based upon a [G]enerative [P]retrained [T]ransformer model \cite{OpenAi,vaswani2017attention, ouyang2022training}. ChatGPT was trained on a huge sample of human text (notably the `Common Crawl' dataset) \cite{NEURIPS2020_1457c0d6} and works by predicting certain combinations of words within a given context. In March 2023, OpenAI released ChatGPT-4 \cite{openai2023gpt4} which is capable of delivering significantly superior performance over its GPT3.5 predecessor. 
 
In July 2023, OpenAI released a beta version of Code Interpreter \cite{CodeInterpreter},  a ground-breaking plugin that increases ChatGPT's capability and potential applications by writing and running Python code, performing mathematical operations, analysing data, and providing graphical visualisations within the standard ChatGPT user interface. Additionally, it allows users to upload data in a number of formats, read files, summarise content, and carry out code-based data analysis, which was previously not feasible using the standard version of ChatGPT3.5 or ChatGPT4. At the time of writing, ChatGPT Code Interpreter is only available for ChatGPT Plus subscription users \cite{ChatGPTPlus}, or ChatGPT Enterprise users \cite{ChatGPTEnterprise}, but will likely be rolled out more widely in the future.

\subsection{AI use in Higher Education}\label{subsec:AIuse}

Given the wide range of potential applications of LLM's, it is not surprising that research has begun to explore these technologies within an educational setting. For example, it has already been demonstrated that ChatGPT4 is able to pass key standardised tests including SAT, LSAT, MCAT and GRE \cite{Insider_tests2023}, as well as the Advance Placement exams in high schools, and Medical and MBA qualifier tests \cite{Kung_medical2023,WhartonMBA2023}. This impressive performance provides additional evidence of ChatGPT's capacity to `understand' and generate responses to complex conceptual questions. 

While AI's capabilities have naturally raised concerns among some school and university teachers regarding academic integrity, there has also been growing interest in how AI might be used to support teaching and learning, although it is still unclear how best to achieve this.  In order to address some of these issues, several policy reports have been commissioned to provide useful guidelines and strategies on how to incorporate AI within the educational system \cite{AI_use_Gov2023, Unesco2023, UKGov2023}. Even though STEM teaching is yet to be significantly impacted by the early ChatGPT models, the advanced features of new incoming AI models and plug-ins can easily be adopted by students to solve complex science and engineering problems \cite{Vasconcelos2023}, and it is therefore essential for STEM educators to critically reflect upon their teaching and assessment practices.  

As one of the key domains in STEM education, Physics presents a compelling arena for testing the capabilities of LLM's. The combination of mathematical equations, data analysis, problem-solving techniques, and sophisticated conceptual understanding makes it an ideal platform for a detailed exploration of the advantages, disadvantages, and unrealised potential of LLMs as learning tools \cite{Kortemeyer_2023, west2023ai, west2023advances,ChatGPTinPER_Bitzenbauer,Dahlkemper2023,Socrates2023,DeathofPhysicsessay_Yeadon2023,AISchool}. 

In one of the first studies to explore AI in a physics context, Kortemeyer \cite{Kortemeyer_2023} evaluated ChatGPT-3.5 using the material of a calculus-based physics course, and found that, despite exhibiting many of the misconceptions and mistakes of a beginner learner, ChatGPT-3.5 was able to pass the course. A similar study by West\cite{west2023ai} revealed the ability of ChatGPT to answer conceptual physics questions about kinematics and Newtonian dynamics. West found that, in some cases, ChatGPT-3.5 could match or even exceed the median performance of a university student who had completed one semester of college physics. In follow-up research conducted by West, it was shown that ChatGPT-4 performed almost as well as an experienced physicist on standard mechanics topics, indicating a significant improvement over GPT-3.5 \cite{west2023advances}.



While some promising early research has been conducted in the area of physics education, it is still unclear what impact AI will have on traditional models of assessment \cite{APcheating,Cotton2023_cheating}, or how AI can be optimally utilised to foster critical thinking within a university setting.
Although AI platforms have their uses, these are yet to be fully integrated into schools and universities, and `natural' intelligence is still considered the best approach for developing productive study habits, workflows, and mathematical skills \cite{AISchool}. Oss \cite{AISchool} suggests that using AI technologies in addition to traditional teaching methods is preferable to simply replacing them. According to Oss, AI should be used as \textquote{objects-to-think-with}, promoting creativity, critical and reflective thinking, and concept comprehension when used in conjunction with well-designed learning activities \cite{Vasconcelos2023}. 

It has been well documented \cite{Socrates2023} that ChatGPT is  prone to errors and misinformation. Anybody who has interacted with ChatGPT for a considerable amount of time will have noticed that the quality and detail of the questions asked and subsequent responses are often correlated. For a user to ensure a high quality output, it is essential to adopt an iterative approach to questioning, whereby the user learns from the preceding question and response in order to improve and clarify the questions posed. While experts may be able to confidently apply such reasoning behaviours when using ChatGPT, it is still unclear whether novice learners are able to recognise flaws, contradictions, and misinformation generated by ChatGPT. When integrating ChatGPT into an educational setting, it is therefore crucial to thoughtfully consider how to cultivate critical thinking amongst students.

\subsection{Brief history on Physics Labs}\label{subsec:physicslabs}

Instructional laboratory courses represent core components of most undergraduate STEM degree programmes, helping students to develop investigative and practical skills, as well as providing opportunities to develop professional research and report-writing skills. Furthermore, effective laboratory instruction has the potential to foster expert-like attitudes and behaviors \cite{Wilcox2018}. They empower students to exercise critical judgement, modify experiments as needed \cite{HolmesLewandowski2020}, and instill a sense of agency and ownership over their research projects \cite{Kalenderetal2021}. 

However, current lab curricula, often characterized by rigid structures and content-reinforcing practices, may fall short in providing these valuable learning experiences \cite{PCAST, smith2020expectations,Holmes2020}.  The prevalent ``cookbook" approach to lab education, in which students follow instructions without necessarily questioning or analysing the underlying process, fails to foster positive attitudes and genuine science practises \cite{HolmesLewandowski2020}. This type of instruction may produce STEM graduates with inadequate skills who lose out on important chances to become experts in the design, planning, and analysis of experiments. 

In light of the evolving educational landscape and the dynamic skill requirements of the job market, traditional approaches, especially in laboratory settings, need to be rethought. This reevaluation becomes particularly urgent considering the rapidly changing needs of education and industry as a result of the emergence of AI technologies like ChatGPT. 

While the role of AI in science teaching has largely focused on assessments \cite{west2023ai,Kortemeyer_2023}, numerous questions remain unanswered regarding the role of AI in science laboratory settings. This paper aims to bridge this gap by exploring how Code Interpreter can be utilised for interpreting and completing an introductory physics lab activity. We also examine the relationship between the level of detail in prompts and quality of output. The subsequent section outlines our research questions regarding the integration of AI within the framework of a conventional, ``cookbook style'' laboratory activity. 

\section{Research Questions}\label{sec:RQs}

In this paper, we address the following three research questions: 

\begin{enumerate}

\item How effectively can ChatGPT Code Interpreter generate, analyse, and interpret data relating to an introductory laboratory activity?
\item What strategies can be implemented to optimise effective communication with ChatGPT during laboratory tasks to ensure high quality, reliable output?  
\item What are the implications of AI tools like ChatGPT Code Interpreter on introductory laboratory courses?
\end{enumerate}

Our goal is to contribute to the ongoing dialogue concerning the role of AI in education, specifically within the context of laboratory science education. By exploring the potential benefits and challenges of using Code Interpreter in laboratory activities, we provide valuable insights for educators, researchers, and educational policymakers adapting to the AI transformation in education.

\section{Method}\label{sec:method}

\subsection{Lab Context}\label{subsec:labcontex}

In this study, we looked at an introductory physics lab activity - \textit{Spring Constant}. This typical traditional lab activity is routinely administered in Semester 1 of Year 1 amongst Physics undergraduate students in many physics departments including our institution. The main goal of the activity is for students to determine the spring constant of a helical spring using two different approaches. The first approach involves suspending masses from the spring, measuring the extension, plotting a mass vs extension graph, then using the gradient to determine a value for the spring constant. The second approach involves displacing a mass on a spring and measuring the period of oscillation, then plotting a period-squared versus mass graph and using the gradient to determine a value for the spring constant.

\begin{figure}[tp]
    \centering
     \setlength{\fboxrule}{0.3pt}  
    \setlength{\fboxsep}{3pt}   
    \fbox{\includegraphics [scale=0.24]{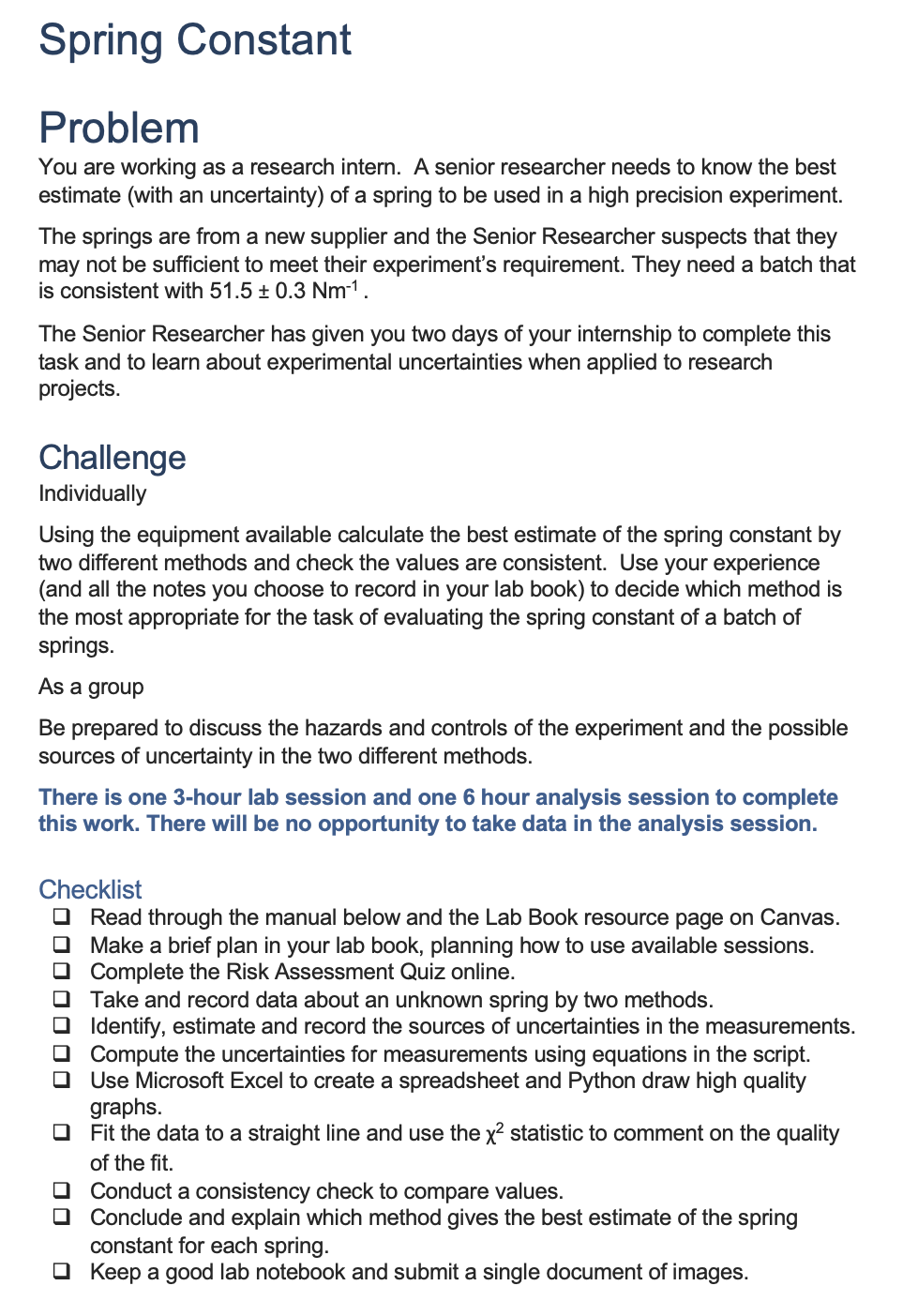}}
    \captionsetup{justification=raggedright, singlelinecheck=false}
    \caption{Spring Constant Lab Activity sheet (student version) which includes purpose of the activity, duration, and a list of step-by-step tasks to complete. The design of this lab activity is considered traditional ``cook-book style" \cite{Wieman2015cognitivelabtask}.}
    \label{fig:Springconstantactivity}
\end{figure}

This activity aims to test students' ability to complete a simple experiment, propagate uncertainties, perform a chi-squared fit of the data, and determine an empirical value for the spring constant, along with an associated uncertainty. The format of lab teaching in this institution is largely traditional or ``cookbook style'' as an analogy where tasks and instructions are presented as a checklist (see Fig \ref{fig:Springconstantactivity}) for students to work through sequentially: collect data, analyse, compare with the true value, and discuss how the procedure could be improved. The expected time to complete this experiment is 9 hours in total for collecting and analysing data and writing results. Students work individually. Instructors and lab demonstrators (graduate students) are available for students to ask questions and help them with the experimental set-up. 

\subsection{Using ChatGPT-4 Advance Data Analysis}

In this study, we considered three separate `Approaches' that made use of Code Interpreter to complete the Spring Constant lab activity.  In all cases, we began by uploading the Lab Activity Microsoft Word document (see supplementary material) using the ChatGPT-4 Code Interpreter interface. In the first two approaches, we asked ChatGPT Code Interpreter to generate simulated data for the experiment through a series of prompts. After Approach 1 (basic prompting), we completed several more attempts to understand the underlying mechanism of Code Interpreter, and how to calibrate our prompts before moving on to Approach 2 (detailed prompting). In Approach 3, we uploaded real experimental data in the form of an Mircosoft Excel spreadsheet, and asked ChatGPT to use this data to complete the activity. 

\section{Results I: ChatGPT Prompts and Responses}\label{sec:results1}

In what follows, we describe all three Approaches used to complete the Spring Constant lab Activity. We describe the key interactions with ChatGPT by highlighting a selection of prompts and responses. In Sec \ref{sec:results2}, we will discuss in more detail the Python code that was written by Code Interpreter to perform the data analysis. You can find the entire conversation for each Approach, along with the Python code generated for each Approach, in the Supplementary documents.

\subsection{Approach 1: Simple Interaction with ChatGPT}\label{subsec:results1Approach1}

After uploading the Lab Activity Word document, we asked ChatGPT: \textit{``Can you work on this activity and answer the questions?"}. ChatGPT responded by providing a concise summary of the activity (Figure \ref{fig:Springconstantsummary}). ChatGPT correctly identified two methods to determine the spring constant. It then laid out a road map for each approach, covering everything from data collection to data fitting, and noted that the $\chi^2$ statistic can be used to comment on the quality of the fit. ChatGPT pointed out that \textit{``conducting the actual experiment and recording physical measurements will have to be done by you, as I am an AI and can't interact with the world"}. At this point, we prompted ChatGPT to generate model data: 

\begin{figure}[tp]
    \centering
    \setlength{\fboxrule}{0.3pt}  
    \setlength{\fboxsep}{0pt}   
    \fbox{\includegraphics [scale=0.41]{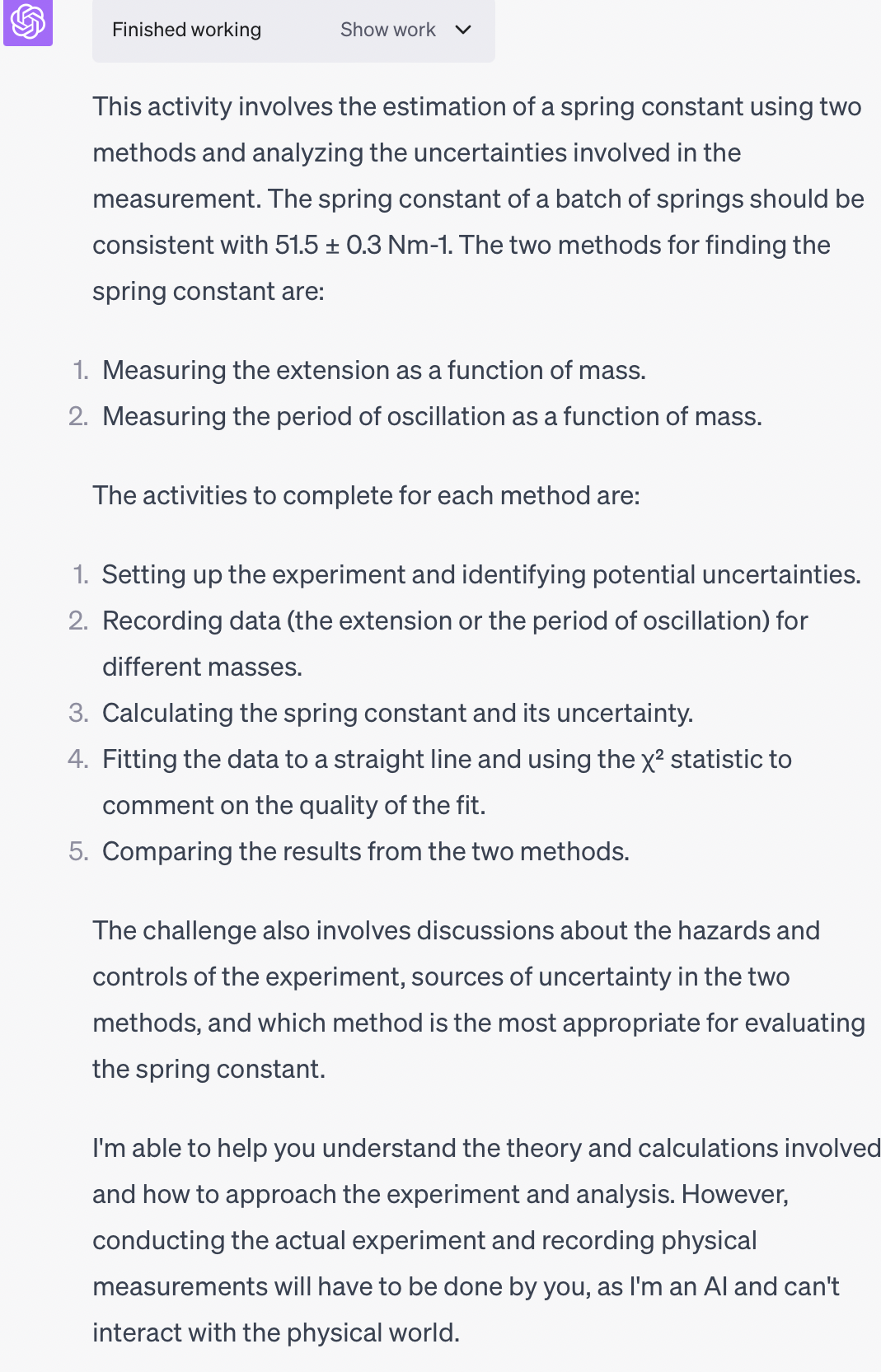}}
    \captionsetup{justification=raggedright, singlelinecheck=false}
    \caption{The response of ChatGPT after uploading the Spring Constant Activity during Approach 1. After successfully processing the Word document, ChatGPT presented a summary describing the activity, discussion points, and the capabilities and limitations of ChatGPT to complete this activity. }
    \label{fig:Springconstantsummary}
\end{figure}

\begin{figure*}[ht!]
    \centering
    \setlength{\fboxrule}{0.3pt}
    \setlength{\fboxsep}{0pt} 
    \fbox{\includegraphics[scale=0.60]{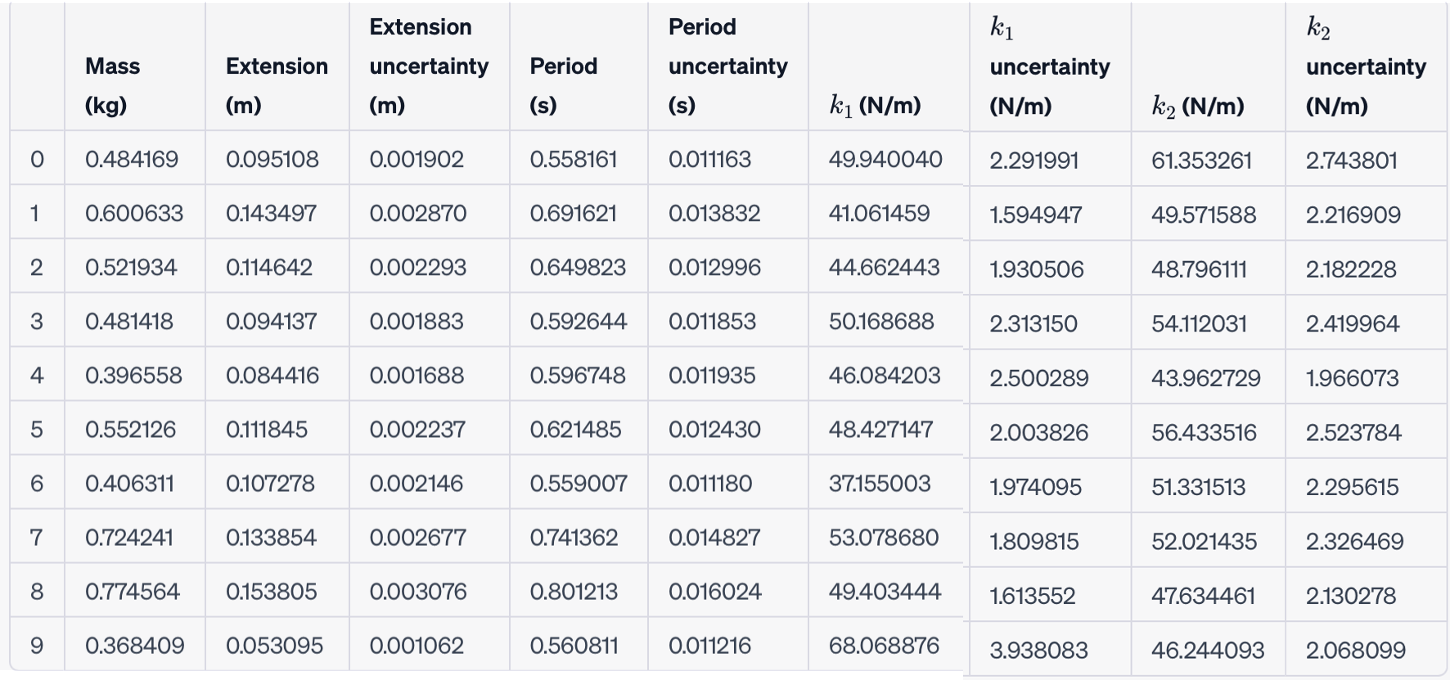}}
     \captionsetup{justification=raggedright, singlelinecheck=false}
    \caption{Output table for Approach 1 showing the mass, extension, and period values, as well as the calculated spring constant values and associated uncertainties for Method 1 and 2, respectively.}
    \label{fig:A}
\end{figure*}

\begin{shadedquotebox}
    {\sffamily\slshape{\textbf{User:} ``Can you generate some model data with associated measurement uncertainties (at least 10 values)? and present the data in a table''}} 
\end{shadedquotebox}

Figure \ref{fig:A} includes the simulated values generated by ChatGPT for different masses, the spring extension, and period of oscillation. The table also provides uncertainty values for \textit{Extension} and \textit{Period}. It is clear from Figure \ref{fig:A} that ChatGPT assigned a 2\% uncertainty to the mass and time period values. In this initial Approach, we deliberately did not ask ChatGPT to explain or justify the uncertainty values, as we wanted to  mimic the behavior of a novice learner. 

We followed up by asking, \textit{``can you now do the activity based on the data above?''}. The calculated values of the spring constant for each method, along with the associated uncertainties, are shown in \ref{fig:A}. It is worth noting that ChatGPT did not fit the data to a straight line and plot a graph, rather opting to calculate the spring constant for each individual set of measurements. 

After prompting ChatGPT to compare the two methods, ChatGPT noted that  \textit{``Method 1 generally results in smaller uncertainties compared to Method 2"} and that \textit{``this could make Method 1 a more reliable method"} and that the accuracy and precision of the measurements could be improved by, \textit{``using more accurate measuring equipment"}, or by, \textit{``taking more measurements to average out random errors"}, comments reminiscent of a novice learner. 

In this first Approach, we did not ask for further explanations, but instead proceeded by asking ChatGPT to generate a \LaTeX{} lab report. A copy of the lab report can be found in the supplementary materials.


\subsection{Approach 2: Including detailed prompts}\label{subsec:results1Approach2}

In Approach 2, we decided to include more detail in our prompts and focus on graph plotting, experimental uncertainties, and error propagation. Whereas in the first Approach we began by uploading the activity and asking, \textit{``can you work on this activity and answer the questions?"}, we decided in the second Approach to change the phrasing to, \textit{``I would like you to read and interpret this activity"}. The hope was that by asking ChatGPT to first \textit{interpret} the activity, a higher quality output would be generated in the conversation. 

We decided to initially focus on \textbf{Method 1} and provide a reasonably detailed prompt that outlined the general features of the approach required to determine the spring constant:

\begin{shadedquotebox}
    {\sffamily\slshape{\textbf{User:} ``Ok, let's start with Method 1. I would like you to pretend you are a human completing this experiment. I would like you to simulate realistic experimental data, as well as specify the relevant measurement uncertainties, and plot a graph with error bars, then perform a chi-squared fit, and determine the reduced chi-squared statistic. You should then use the graph to determine k. You should use the uncertainty in the gradient to determine the uncertainty in k by propagating the uncertainties correctly."}}
\end{shadedquotebox}

\begin{figure}[tp]
    \centering
    \includegraphics [scale=0.17]{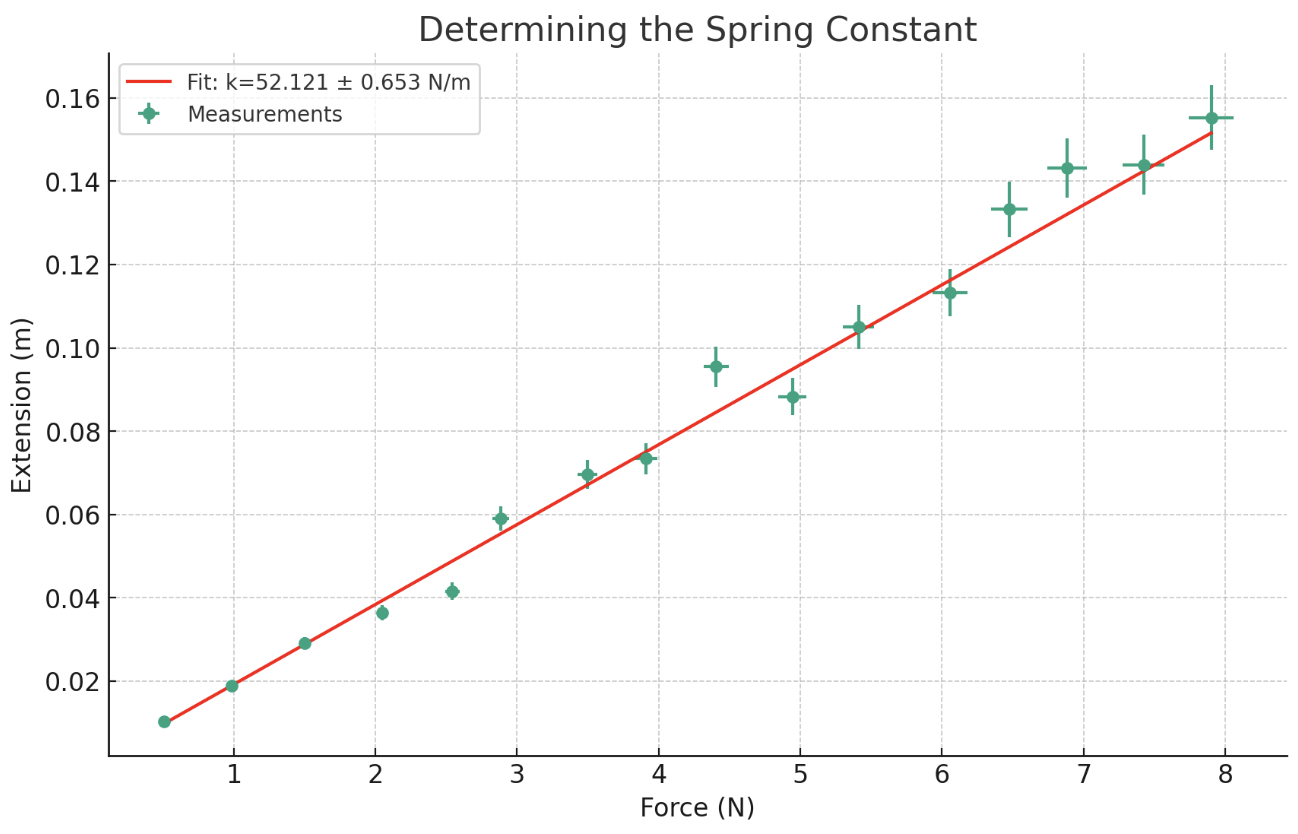}
      \captionsetup{justification=raggedright, singlelinecheck=false}
    \caption{Plot showing Extension ($x$) versus Force ($F$) for Method 1 of the Spring Constant Activity (Approach 2).}
    \label{fig:GraphOutput1}
\end{figure}

When reading the initial response, we noticed that ChatGPT had suggested a range of forces between $1N$ and $10N$. However, since the instructions explicitly stated a maximum mass value of $800g$, we provided an additional prompt: \textit{``the instructions tell you that the masses vary from $50g$ to $800g$''}. ChatGPT then proceeded to write the necessary Python code to simulate and plot the data. The graphical output can be seen in Figure \ref{fig:GraphOutput1}.

ChatGPT reported a spring constant of $52.12\pm 0.65{} N/m$, and a reduced chi-squared statistic of 2.01, suggesting that \textit{``the fit is reasonable, but the model might not perfectly represent the data, or the uncertainties might be slightly underestimated.''}. ChatGPT also noted that \textit{``the estimated spring constant is consistent with the true value used to generate the data, within the uncertainty"}.

At this stage, we decided to probe ChatGPT to understand how it had determined the uncertainty in the spring constant from the model fit data, given that we had not provided any specific details in our prompt. We were keen to see whether ChatGPT had correctly propagated the uncertainty in the gradient to determine the uncertainty in the spring constant. We began by asking:

\begin{shadedquotebox}
    {\sffamily\slshape{\textbf{User:} ``Can you explain, step by step, how you used the fit and the uncertainty in the gradient to determine the uncertainty in the spring constant according to method 1?"}} 
\end{shadedquotebox}

The response began by stating which variables were plotted on the $x$ and $y$ axes. ChatGPT correctly identified that the spring constant would be equal to: $k=1/G$ where $G$ is the gradient of the graph, and recognised that the magnitude of the uncertainty in the spring constant would be $\Delta k = \frac{1}{G^2} \Delta G$ where $\Delta G$ is the uncertainty in the gradient (determined from the line fit). 

For \textbf{Method 2}, we decided to build on the responses provided by ChatGPT during \textbf{Method 1} and provide additional detail, including step by step instructions of how the uncertainty in the spring constant could be determined using the uncertainty in the gradient of the graph:

\begin{shadedquotebox}
    {\sffamily\slshape{\textbf{User:} ``we want to determine the spring constant by an alternative method that involves timing the period of oscillation of a mass attached to a spring as it undergoes simple harmonic motion. In this experiment, we will consider 100g, 200g, ..., 800g. For each mass, we will record the time period of 20 complete oscillations. For each mass, we will repeat this process 5 times, and then calculate the mean value corresponding to a single oscillation. We will then calculate the standard deviation of the period, and the standard error on the mean. We will do this for each mass. We will then plot a graph with $T^2$ on the y-axis and m on the x-axis. The gradient of this graph will be equal to $4\pi^2/k$, and therefore the spring constant will be equal to: $k = 4\pi^2$/gradient.  The uncertainty in the gradient will be equal to: $\Delta k = |{\partial k}/ {\partial G}| * \Delta G = 4 \pi^2 / G^2 * \Delta G$ where $\Delta G$ is the uncertainty in the gradient, and $G$ is the gradient."}} 
\end{shadedquotebox}

We provided further detailed instructions relating to the nature of the simulated data, the plot, and the propagation of uncertainties: 

\begin{shadedquotebox}
    {\sffamily\slshape{\textbf{User:} ``With this in mind, I would like you to create a table of model data with the following columns: $m, \Delta m, T_{mean}, \Delta T, T^2, \Delta T^2$. The mass values should vary from 50g to 800g. We will assume that the uncertainty in the mass value  $\Delta m$ is $5\%$ of the stated mass value. For the time period measurements,  assume that the biggest source of uncertainty is determining the start and end point of the oscillation and the random effect this will have on the measured time periods. The data should feel realistic and subject to random noise. The aim should be that when the data is plotted, the points do not lie exactly on the line, and that the error bars representing $\Delta m$ and $\Delta T^2$ should be clearly visible on the graph. The data should be consistent  with a spring constant of $51.5 N/m$ within the limits of the stated uncertainty. Plot a graph with $T^2$  on the y-axis, and mass on the x-axis, along with error bars represented by $\Delta T^2$ and $\Delta m$. You should perform a chi-squared fit of the data and determine the reduced chi-squared statistic, which should be between 0.5 and 4, and the spring constant. By using the uncertainty in the gradient,  you should calculate the  uncertainty in the spring constant which will be $\Delta k = 4 \pi^2 / G^2 * \Delta G$ where $\Delta G$ is the uncertainty in the gradient, and $G$ is the gradient"}} 
\end{shadedquotebox}

\begin{figure}[tp]
    \centering
    \includegraphics [scale=0.17]{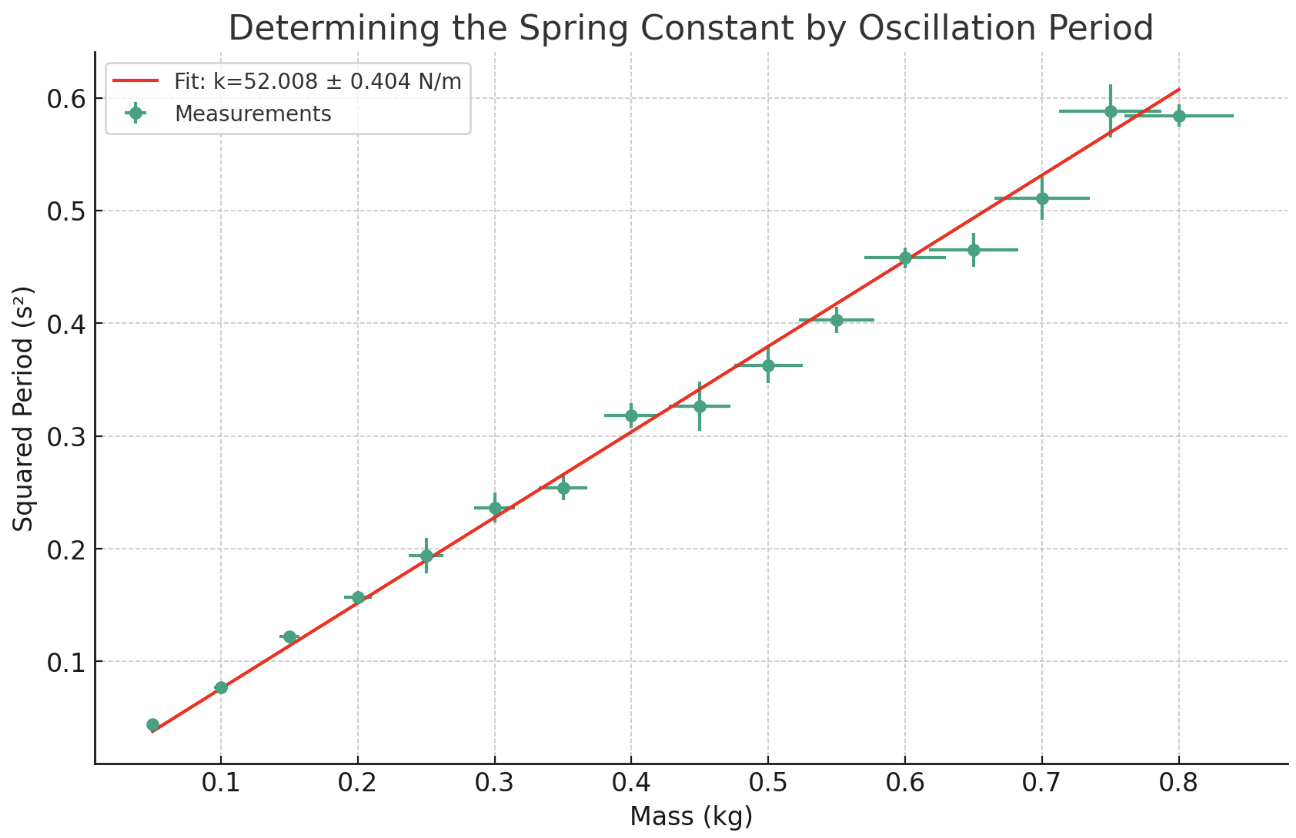}
      \captionsetup{justification=raggedright, singlelinecheck=false}
    \caption{Plot with squared values of period ($T^2$) on the y-axis versus mass ($m$) values on the x-axis for Method 2 of the Spring Constant Activity (Approach 2). Error bars included for both variables.}
    \label{fig:Method2Graph}
\end{figure}

 The plotted data and line fit can be seen in Figure \ref{fig:Method2Graph}. The method used by ChatGPT to generate the data points and uncertainty values for time period will be discussed in the next section. ChatGPT reported a spring constant of $51.37 \pm 0.66 N/m$, and a reduced chi-squared statistic of $1.61$, suggesting a reasonably good fit. ChatGPT commented that the spring constant value was consistent with the `true' spring constant value of $51.5 N/m$, within the limits of the stated uncertainty. 

 ChatGPT further noted that Method 1 \textit{``requires accurate measurements of small extensions, which can be challenging"}, whereas Method 2 requires the timing of oscillations, which can be challenging for fast oscillations and \textit{``calculating the period from the timing of multiple oscillations)"}. We ended Approach 2 by asking ChatGPT to write a \LaTeX{} lab report (supplementary materials).

\subsection{Approach 3: Uploading experimental data}\label{subsec:results1Approach3}

 In Approach 3, we explored ChatGPT Code Interpreters ability to complete the activity using `real' experimental data that had been stored in a Microsoft Excel spreadsheet. This is likely the most common way in which students will use ChatGPT Code Interpreter. After uploading the Spring Constant Lab activity and asking ChatGPT to \textit{``please complete the activity"}, the excel file was uploaded and ChatGPT was prompted to \textit{``read both sheets of the data"}. We decided to explicitly prompt ChatGPT to perform an Orthogonal Distance Regression (ODR) fit to ensure that both $x$ and $y$ errors were included when performing the fit:

\begin{shadedquotebox}
    {\sffamily\slshape{\textbf{User:} ``Ok, can you now use the data from Sheet 1. There are four columns, the data in column 1 and 2 represents the y-data and y-error respectively. The data in columns 3 and 4 represents the x-data and x-error respectively. Convert into SI units. Perform an ODR fit. Then use the fit data to calculate the spring constant and uncertainty in spring constant. The spring constant will be equal to the gradient multiplied by g. Also state the reduced chi-squared statistic for the plot."}} 
\end{shadedquotebox}

\begin{figure}[tp]
    \centering
    \includegraphics [scale=0.50]{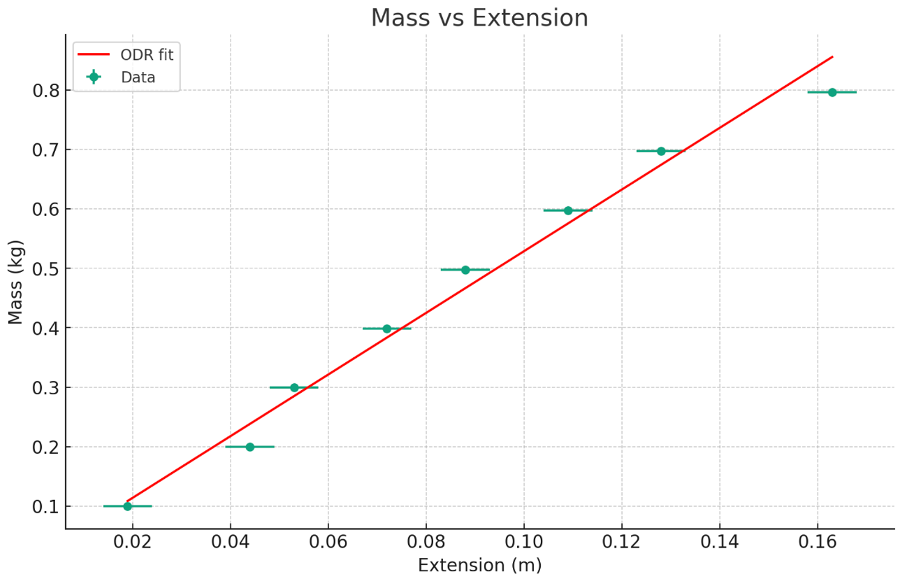}
     \captionsetup{justification=raggedright, singlelinecheck=false}
    \caption{Plot showing mass ($m$) versus extension ($x$) for Method 1 of the Spring Constant Activity (Approach 3). The estimated uncertainty on the mass is too small to see on this plot.}
    \label{fig:3}
\end{figure}

ChatGPT then proceeded to write and run the necessary Python code, plot a graph (Figure \ref{fig:3}), before responding with a value for the spring constant of $50.86 \pm 2.80 N/m$ and a reduced chi-squared statistic of $1.88$. 

Next, we asked ChatGPT to \textit{``explain, step by step, how you determined the uncertainty in the spring constant"}. ChatGPT responded by noting that the spring constant can be determined from the gradient, $G$, of the Mass versus Extension graph as $k=g.G$, where $g$ is the gravitational field strength. ChatGPT noted that the error in the spring constant, $\Delta k$, is given by $\Delta k = |dk/dG|.\Delta G $ where $\Delta G$ refers to the uncertainty in the gradient (extracted from the ODR fit parameters), and since $ k=g.G $, it follows that $\Delta k = |g|.\Delta G$. 

An identical set of instructions was provided to ChatGPT for \textbf{Method 2}, this time referring to the data in Sheet 2. After performing an ODR fit, ChatGPT correctly determined the spring constant from the gradient of a $T^2$ versus $m$ graph, arriving at a value of $65.97 \pm 0.75 N/m$, with a reduced chi-square statistic of 12.32. ChatGPT noted that the value of 12.32 is \textit{``significantly larger than 1, suggesting that the fit is not as good as it could be"}. It then went on to say that this might be due to, \textit{``systematic errors in the experiment"}, but did not elaborate further.

Next, we prompted ChatGPT to explain how the spring constant uncertainty value was determined. ChatGPT correctly identified that the uncertainty in the spring constant is given by $\Delta k = |-4\pi^2/G^2|.\Delta G$ where $G$ is the gradient of the $T^2$ vs $m$ graph. Finally, we asked ChatGPT to produce a \LaTeX lab report using an identical set of prompts to those used in Approach 2 (supplementary materials).

\section{Results II: Python Code}\label{sec:results2}
One of the primary tasks in this research study was to establish whether ChatGPT Code Interpreter was capable of generating and analysing data relating to the uploaded Spring Constant Activity. In all three \textit{Approaches}, Code Interpreter wrote Python code in response to our prompts. Here we will describe and discuss the Python code that was generated. Table \ref{table:comparison2} summarises the features of data generation and analysis by comparing \textit{Approach 1} and \textit{Approach 2}.

\subsection{Approach 1}

\begin{table*}[tp]
    \centering
    \small 
    \arrayrulecolor{gray!75} 
    \renewcommand{\arraystretch}{1.5} 
    \setlength{\arrayrulewidth}{0.8pt} 
    \begin{tabular}{|p{0.25\textwidth}|p{0.35\textwidth}|p{0.35\textwidth}|} 
        \hline
        \rowcolor{Gray}\textbf{Aspect} & \textbf{Approach 1} & \textbf{Approach 2} \\ 
        \hline
        Generating mass values & A random uniform distribution was used to generate 10 masses between 0.1kg and 0.8kg. & A linearly spaced method to generate 16 masses between 0.05kg and 0.8kg. \\
        \hline
        Generating ‘true’ extension and period values & Forces calculated from masses by multiplying with \( g \), and then used Hooke’s law to generate the true extension for these forces using $x=mg/k$. True time periods calculated using $T=2\pi \sqrt{m/k}$. & Forces calculated from masses by multiplying with \( g \), and then used Hooke’s law to generate the true extension for these forces using $x=mg/k$. True time periods calculated using $T=2\pi \sqrt{m/k}$. \\
        \hline
        Adding noise to simulate random measurement error & The standard deviation of the noise was fixed at 0.02 irrespective of the value of extension, and noise was added uniformly to all of the data points. & The standard deviation of the noise was variable, and was given by the product of the relative uncertainty and the `true` value (of force, extension, or time period). The magnitude of the noise was proportional to the magnitude of the data point (larger forces/extensions/periods had larger absolute noise). \\
        \hline
        Errors & The error was assumed to be \( 2\% \) of both mass, extension and period measurements. & \( 2\% \) for the force measurements, \( 5\% \) for extension and period measurements. For the period measurements, the error for the period was taken as the standard error on the mean of five repeat measurements. \\
        \hline
    \end{tabular}
    \captionsetup{justification=raggedright, singlelinecheck=false}
    \caption{Describing the four key aspects of the data generation process using ChatGPT Code Interpreter. A comparison of Approach 1 and 2.}
    \label{table:comparison2}
\end{table*}

After prompting ChatGPT to \textit{``generate some model data"},  ChatGPT began by extracting the `true' value of the spring constant from the Spring Constant lab activity. After importing the necessary \texttt{numpy} python library, ChatGPT generated an array of 10 random masses between 0.1kg and 0.8kg:
\begin{mdframed}[linecolor=black, backgroundcolor=gray!10, innerleftmargin=8pt, innerrightmargin=0pt, innertopmargin=0pt, innerbottommargin=0pt]
\begin{lstlisting}
%k_true = 51.5 
g = 9.81  
np.random.seed(0)  
masses = np.random.uniform(0.1, 0.8, 10)
\end{lstlisting}
\end{mdframed}

 ChatGPT used \texttt{k\_true} (shaded red), along with the randomly generated masses, to reverse engineer values for the extension and time periods using the equations $x=mg/k$ and $T=2\pi \sqrt{m/k}$ respectively:

\begin{mdframed}[linecolor=black, backgroundcolor=gray!10, innerleftmargin=8pt, innerrightmargin=0pt, innertopmargin=0pt, innerbottommargin=0pt]
\begin{lstlisting}
extensions = masses * g / k_true + np.random.normal(0, 0.02, 10) 
periods = 2 * np.pi * np.sqrt(masses / k_true) + np.random.normal(0, 0.02, 10)  
\end{lstlisting}
\end{mdframed}

We see that random noise from a normal distribution was added to these calculated values to simulate measurement error. A constant percentage uncertainty of $2\%$ for each individual `measurement' was assumed, and this was used as the standard deviation of the normal distribution of noise that was added to the `true' values. 

For \textbf{Method 1}, the spring constant was calculated as $k=F/x$, and the uncertainty in each spring constant value, $\Delta k$, was determined by propagating the uncertainties in the force, $\Delta F$, and the uncertainty in the extension, $\Delta x$, using the error propagation equation, $\Delta k = k\sqrt{(\Delta F/F)^2+(\Delta x/x)^2}$, as seen below:

\begin{mdframed}[linecolor=black, backgroundcolor=gray!10, innerleftmargin=8pt, innerrightmargin=0pt, innertopmargin=0pt, innerbottommargin=0pt]
\begin{lstlisting}
forces = data['Mass (kg)'] * g
k1 = forces / data['Extension (m)']
%force_uncertainties = g * 0.02  
k1_uncertainties = k1 * np.sqrt((force_uncertainties / forces)**2 + (data['Extension uncertainty (m)'] / data['Extension (m)'])**2)
\end{lstlisting}
\end{mdframed}

We note that the uncertainty in the force, $\Delta F$, was \textit{incorrectly} calculated as $\Delta F = g \times 0.02$ (highlighted in red in the code above), which would imply that each force value had a constant uncertainty equal to $2 \%$ of the gravitational field strength. However, the correct equation for the uncertainty in the force should have been $\Delta F = F\sqrt{(\Delta m/m)^2+(\Delta g/g)^2}$, which reduces to $\Delta F = F(\Delta m/m)$ if we assume that the uncertainty in the gravitational field strength is negligible. Since the fractional uncertainty in the mass was assumed to be $2\%$, the uncertainty in the force should have been calculated as $\Delta F = F \times 0.02 = m g \times 0.02$, and so we see that the expression provided by ChatGPT was missing a factor of $m$.

For \textbf{Method 2}, the spring constant was correctly calculated as $k=4 \pi^2 m/T^2$, and the uncertainty in the spring constant was correctly determined by propagating the uncertainties in mass, $\Delta m$, and time period, $\Delta T$, as seen in the code snippet below:

\begin{mdframed}[linecolor=black, backgroundcolor=gray!10, innerleftmargin=8pt, innerrightmargin=0pt, innertopmargin=5pt, innerbottommargin=5pt]
\begin{lstlisting}
k2 = 4 * np.pi**2 * data['Mass (kg)'] / data['Period (s)']**2
mass_uncertainties = data['Mass (kg)'] * 0.02 
%k2_uncertainties = k2 * np.sqrt((mass_uncertainties / data['Mass (kg)'])**2 + (2 * data['Period uncertainty (s)'] / data['Period (s)'])**2)
\end{lstlisting}
\end{mdframed}

It might not be immediately obvious that the red-shaded expression above is the correct error propagation equation, since $k$ is a function of $m$ and $T^2$, and therefore the `correct' error propagation equation should be $\Delta k = k\sqrt{(\Delta m/m)^2+(\Delta T^2/T^2)^2}$. However, this is equivalent to the expression generated by ChatGPT since $\Delta (T^2)=2T\Delta T$, and therefore the relative uncertainty in $T^2$ is given by $\Delta (T^2)/T^2 = 2(\Delta T/T)$. Substituting this into the error propagation equation shows the equivalence of the two expressions. 

 It is interesting to note that during Approach 1, ChatGPT opted to calculate the spring constant separately for each pair of `measured' masses and time periods (see Figure \ref{fig:A}), rather than taking the standard graphical approach and plotting a graph with $T^2$ on the y-axis and $m$ on the x-axis, performing a line fit, and then using the gradient to determine the spring constant. Although we do not know the reasons for ChatGPT's `choices' in Approach 1, we will see that by changing the prompts in Approach 2, the methods used by Code Interpreter will also change.

\subsection{Approach 2}

In Approach 2, we included more detail in our prompts with a focus on graph plotting, experimental uncertainties, and error propagation. This resulted in a more sophisticated approach. For \textbf{Method 1}, ChatGPT used the `true' spring constant, along with 16 equally spaced mass values between 0.05kg and 0.8kg, to reverse engineer the `true' extension values using Hooke's Law. A relative uncertainty of $2\%$ was assumed for the force measurements, and $5\%$ for the extension measurements. The `true' force and extension values were then used to generate \texttt{measured\_forces} and \texttt{measured\_extensions}: 

\begin{mdframed}[linecolor=black, backgroundcolor=gray!10, innerleftmargin=8pt, innerrightmargin=0pt, innertopmargin=0pt, innerbottommargin=0pt]
\begin{lstlisting}
def hooke_law(force, k):
    return force / k
np.random.seed(0)  
true_extensions = hooke_law(forces, true_k)
%measured_forces = np.random.normal(forces, force_uncertainty * forces)
%measured_extensions = np.random.normal(true_extensions, extension_uncertainty * true_extensions)
\end{lstlisting}
\end{mdframed}

We see from the red-highlighted code segments above, \texttt{measured\_forces} was drawn from a normal distribution centered at the `true' force, with a standard deviation of $2\%$ of the true force. A similar approach was used to generate \texttt{measured\_extensions}, this time with a standard deviation of $5\%$ of the true extension. Since the standard deviation of the noise scales with the size of the force and extension, it follows that when the data is plotted, the points corresponding to larger force and extension values are scattered further from the fitted line than the points corresponding to smaller extension/force values. This heteroscedasticity in the data is precisely what we observe in Figure \ref{fig:GraphOutput1}. Further discussion of this point can be found in the Appendix.

Next, having explicitly prompted ChatGPT to perform a line fit on the simulated data, ChatGPT used the \texttt{curve\_fit} function from the \texttt{SciPy} library to fit the \texttt{hooke\_law} function to the synthetic data generated earlier in the code:
\begin{mdframed}[linecolor=black, backgroundcolor=gray!10, innerleftmargin=8pt, innerrightmargin=0pt, innertopmargin=5pt, innerbottommargin=5pt]
\begin{lstlisting}
popt, pcov = curve_fit(hooke_law, measured_forces, measured_extensions, %sigma=extension_errors, absolute_sigma=True)
fitted_k = popt[0]
fitted_k_error = np.sqrt(pcov[0][0])
\end{lstlisting}
\end{mdframed}

We note from the red-highlighted code above that despite ChatGPT generating synthetic data that included uncertainties in both force  and extension, the \texttt{curve\_fit} function used for the line-fitting only makes use of the $y$-error's (extension), and neglects to factor in the $x$-errors (force). We see from the code snippet that the spring constant $k$ is obtained from \texttt{popt[0]} and the uncertainty in $k$ is the square root of the first diagonal element of the covariance matrix \texttt{np.sqrt(pcov[0][0])}. 

Next, following our detailed prompt, ChatGPT proceeded to calculate the reduced chi-squared statistic. The first step was to calculate the \texttt{fitted\_extensions} using the $k$ value from the line-fit along with the \texttt{hooke\_law} function. We see from the red-highlighted code below that the y-residuals (difference between `measured' and `fitted' extensions) were determined and used to calculate the chi-square statistic (which is the sum of the squared standardised residuals). This involved dividing each residual by the corresponding uncertainty \texttt{`extension\_errors'} to standardise it, before squaring and then summing all the standardised residuals. The reduced chi-square statistic was then calculated by dividing the chi-square statistic by the number of degrees of freedom, which is the number of data points, \texttt{len(measured\_forces)}, minus the number of fitted parameters, \texttt{len(popt)}:
\begin{mdframed}[linecolor=black, backgroundcolor=gray!10, innerleftmargin=8pt, innerrightmargin=0pt, innertopmargin=0pt, innerbottommargin=0pt]
\begin{lstlisting}
fitted_extensions = hooke_law(measured_forces, fitted_k)
%residuals = measured_extensions - fitted_extensions
%chi_square = np.sum((residuals / extension_errors)**2)
reduced_chi_square = chi_square / (len(measured_forces) - len(popt))
\end{lstlisting}
\end{mdframed}

 We note that since the line-fit (and hence \texttt{fitted\_k}) and residuals only factored in the extension errors (and not the force errors), the (reduced) chi-square statistic calculated by ChatGPT likewise only factored in the extension-errors, though this was not explicitly mentioned in the ChatGPT response. Finally, following our explicit prompt, a graph was plotted (see Figure \ref{fig:GraphOutput1}) using the standard \texttt{matplotlib} Python library.

For \textbf{Method 2}, we prompted ChatGPT to simulate 5 repeated measurements for the time period (see supplementary material). ChatGPT began by using the `true' spring constant value, along with the equation for the period of oscillation of a mass on a spring, $T=2\pi \sqrt{m/k}$, to generate `true' time periods (\texttt{true\_periods}). Then, for each value in \texttt{true\_periods}, the code generated an array of 5 \texttt{measured\_periods} by drawing from a normal distribution with a mean equal to the true period value, and a standard deviation equal to $5\%$ of the true period value (shown in red below):

\begin{mdframed}[linecolor=black, backgroundcolor=gray!10, innerleftmargin=8pt, innerrightmargin=0pt, innertopmargin=0pt, innerbottommargin=0pt]
\begin{lstlisting}
np.random.seed(0) 
true_periods = oscillation_period(masses, true_k)
%measured_periods = np.array([np.random.normal(true_period, 0.05*true_period, 5) for true_period in true_periods])
\end{lstlisting}
\end{mdframed}

It is worth noting that, much like with Method 1, because the standard deviation of the measurement noise was made proportional to the true value, the `measured periods' appeared more scattered around the true value for larger period values. You can clearly observe this heteroscedasticity in the plotted data appearing in the $T^2$ versus $m$ graph in Figure \ref{fig:Method2Graph}. This point is discussed in more detail in the Appendix.

In Approach 2, we explicitly prompted Chat GPT to calculate the mean, standard deviation and standard error using the `repeat measurements', as can be seen in the code chunk below:

\begin{mdframed}[linecolor=black, backgroundcolor=gray!10, innerleftmargin=8pt, innerrightmargin=0pt, innertopmargin=0pt, innerbottommargin=0pt]
\begin{lstlisting}
mean_periods = np.mean(measured_periods, axis=1)
std_periods = np.std(measured_periods, axis=1, ddof=1)
sem_periods = std_periods / np.sqrt(measured_periods.shape[1])
\end{lstlisting}
\end{mdframed}

Next, ChatGPT correctly calculated the squared Periods and the error on the squared Periods, $\Delta T^2$, by propagating the uncertainty in $T$ (the standard error on the mean Time Period) according to the relation $\Delta T^2 = 2T\Delta T$ (highlighted red): 
\begin{mdframed}[linecolor=black, backgroundcolor=gray!10, innerleftmargin=8pt, innerrightmargin=0pt, innertopmargin=0pt, innerbottommargin=0pt]
\begin{lstlisting}
squared_periods = mean_periods**2
%squared_periods_errors = 2 * mean_periods * sem_periods
\end{lstlisting}
\end{mdframed}

The next step was similar to Method 1 where ChatGPT wrote the code to perform the line fit and calculate the spring constant, and its uncertainty, from the model fit parameters:

\begin{mdframed}[linecolor=black, backgroundcolor=gray!10, innerleftmargin=8pt, innerrightmargin=0pt, innertopmargin=0pt, innerbottommargin=0pt]
\begin{lstlisting}
popt, pcov = curve_fit(squared_period_model, masses, squared_periods, sigma=squared_periods_errors, absolute_sigma=True)
fitted_slope = popt[0]
fitted_slope_error = np.sqrt(pcov[0][0])
%fitted_k = 4 * np.pi**2 / fitted_slope
%fitted_k_error = np.abs(4 * np.pi**2 / fitted_slope**2 * fitted_slope_error)
\end{lstlisting}
\end{mdframed}

We note from the red-highlighted code above that the fitted spring constant (\texttt{fitted\_k}) was correctly determined from the gradient (\texttt{fitted\_slope}) according to the relation $k=4 \pi^2/G$ (where $G$ is the gradient).  We also note that the uncertainty in the spring constant was correctly calculated by propagating the uncertainty in gradient, $\Delta G$, using the relation: $\Delta k = |4\pi^2/G^2| \Delta G $. 

Again, similar to Method 1, the chi-square statistic was calculated by taking each residual (difference between the `measured' $T^2$ values and the fitted $T^2$ values), dividing by the corresponding uncertainty $\Delta T^2$, squaring the result, and then summing over all data points. We note that, as with Method 1, the calculation of the chi-square statistic only involved use of the $\Delta T^2$ errors.

Finally, the data was plotted, along with error bars representing $\Delta T^2$ and $\Delta m$ (see Figure \ref{fig:Method2Graph}). We note that error bars representing both $\Delta T^2$ and $\Delta m$ were included in the plot, despite the fact that only the $\Delta T^2$ errors were considered in the least squares line fit. It is interesting to note that, similar to Method 1, even with detailed prompts, ChatGPT opted for a line-fitting procedure that neglected the errors in the mass values. 

For this reason, we decided to change the phrasing of our prompts during Approach 3 and explicitly ask ChatGPT to perform an Orthogonal Distance Regression (ODR) fit to the data, since this type of fit factors in both the $x$-errors and $y$-errors.

\subsection{Approach 3}

In Approach 3, we uploaded real student data to Code Interpreter in the form of an Excel spreadsheet. The spreadsheet consisted of two sheets, one for each method. Each sheet contained four columns corresponding to $x$-values, $x$-errors, $y$-values, and $y$-errors. 

In \textbf{Method 1}, ChatGPT began by loading both sheets from the Excel file into separate dataframes using Python's Pandas module \texttt{read\_excel}:

\begin{mdframed}[linecolor=black, backgroundcolor=gray!10, innerleftmargin=8pt, innerrightmargin=0pt, innertopmargin=0pt, innerbottommargin=0pt]
\begin{lstlisting}
df1 = pd.read_excel(xls, 'Method 1')
df2 = pd.read_excel(xls, 'Method 2')
df1, df2
\end{lstlisting}
\end{mdframed}

Having prompted ChatGPT (see Section\ref{subsec:results1Approach3}) to perform an ODR fit, ChatGPT began by importing the necessary Python libraries and defining the linear model (in red):

\begin{mdframed}[linecolor=black, backgroundcolor=gray!10, innerleftmargin=8pt, innerrightmargin=0pt, innertopmargin=0pt, innerbottommargin=0pt]
\begin{lstlisting}
from scipy.odr import ODR, Model, Data
import numpy as np
% def f(B, x):
   % return B[0]*x + B[1]
\end{lstlisting}
\end{mdframed}

Next, following the prompt to perform an ODR fit, ChatGPT began by assigning data and errors to appropriately named variables:
\begin{mdframed}[linecolor=black, backgroundcolor=gray!10, innerleftmargin=8pt, innerrightmargin=0pt, innertopmargin=0pt, innerbottommargin=0pt]
\begin{lstlisting}
y = df1['mass (g)'].values
y_err = df1['error (g)'].values
x = df1['extension (mm)'].values
x_err = df1['error (mm)'].values
\end{lstlisting}
\end{mdframed}

ChatGPT then proceeded to run the ODR fit by first creating an instance of the \texttt{Model} class provided by \texttt{scipy.odr}, then creating an instance of the \texttt{Data} class which was used to store the data (both $x$ and $y$ variables) and their associated uncertainties (or weights) for the ODR fit. Finally, an instance of the ODR class was created with the data, the model, and an initial estimate of the parameters ( \texttt{B[0]} and \texttt{B[1]} equal to 0). The ODR fit was then run, and the model fit parameters were extracted:

\begin{mdframed}[linecolor=black, backgroundcolor=gray!10, innerleftmargin=8pt, innerrightmargin=0pt, innertopmargin=0pt, innerbottommargin=0pt]
\begin{lstlisting}
linear = Model(f)
mydata = Data(x, y, wd=1./x_err**2, we=1./y_err**2)
myodr = ODR(mydata, linear, beta0=[0., 0.])
output = myodr.run()
B = output.beta
B_err = output.sd_beta
\end{lstlisting}
\end{mdframed}

The model fit parameter, \texttt{B[0]} (representing the slope of the fitted line) and \texttt{B\_err[0]} (representing the error in the slope) were then correctly used to to calculate a value for the spring constant $k$ and the uncertainty in the spring constant, $\Delta k$ (shaded red):

\begin{mdframed}[linecolor=black, backgroundcolor=gray!10, innerleftmargin=8pt, innerrightmargin=0pt, innertopmargin=0pt, innerbottommargin=0pt]
\begin{lstlisting}
g = 9.81  
%k = B[0] * g
%k_err = B_err[0] * g
\end{lstlisting}
\end{mdframed}


The reduced chi-square statistic was then calculated using the \texttt{res\_var} attribute from the ODR output, explicitly accounting for uncertainties in both the $x$ and $y$ variables. This follows from the fact that the ODR chi-square statistic is formed by squaring the weighted orthogonal distances, and then summing over all data points. Since the weighted orthogonal distance factors in uncertainties in both x and y variables, it follows that the same is true for the chi-square statistic. Finally, the data was plotted with error bars, with the fit line overlaid on the plotted data points (Figure \ref{fig:3}). 

\textbf{Method 2:} ChatGPT began by extracting the relevant measured data, along with uncertainties, from Sheet 2 of the Excel file and assigning them to separate variables. ChatGPT then ran the ODR fit, extracted the fit parameters, and used these to calculate $k$, $\Delta k$, and the reduced chi-square statistic. The Python code used was similar to Method 1, although the spring constant and uncertainty in the spring constant had to be calculated from the fit parameters using error propagation:

\begin{mdframed}[linecolor=black, backgroundcolor=gray!10, innerleftmargin=8pt, innerrightmargin=0pt, innertopmargin=0pt, innerbottommargin=0pt]
\begin{lstlisting}
output = myodr.run()
B = output.beta
B_err = output.sd_beta
k = 4 * np.pi**2 / B[0]
k_err = 4 * np.pi**2 * B_err[0] / B[0]**2
chi_square = output.res_var
\end{lstlisting}
\end{mdframed}

We note that ChatGPT correctly calculated $k=4 \pi^2/G$ where $G=B[0]$ represents the slope of the line fit, and $\Delta k = 4 \pi^2 \Delta G / G^2$ where \texttt{B\_err[0]} is equal to the uncertainty in the slope, $\Delta G$. Again, the reduced chi-square statistic was calculated using the \texttt{res\_var} attribute from the ODR output. The data was then plotted with error bars, with the fit line overlaid. 

In summary, ChatGPT was able to use our prompts to correctly perform an ODR fit, calculate the reduced chi-square statistic, and plot the data with an appropriate line fit, producing results identical to those obtained by the researchers using the same data set.

\section{Discussion}

This study has demonstrated that although ChatGPT Code Interpreter was capable of interpreting and completing an introductory laboratory activity, the quality and accuracy of ChatGPT's output varied significantly across three different Approaches, highlighting the importance of prompt detail and specificity when using Code Interpreter for data generation and analysis. As outlined in the results section, we initially prompted ChatGPT to ``work on" the activity and simulate data to complete the task (Approach 1). The initial prompts used were specifically written to mimic a novice ChatGPT user and a novice learner in physics. Expanding on this, the interactions of Approach 2 included more detailed instructions relating to the simulation and analysis of data, and were more representative of expert-level knowledge. Results from Section \ref{sec:results1} and \ref{sec:results2} revealed some changes in ChatGPT's approach to data generation and analysis across Approach 1 and 2. Lastly, Approach 3 made use of `real' student data and involved targeted prompts relating to data analysis that built on Approach 1 and 2. In the subsequent sections, we address each of our research questions by discussing the key insights gained from our interactions with ChatGPT, the quality of the data generated and analysed by ChatGPT, and the possible implications for lab instructors and students. 

\begin{table*}[tp]
    \centering
    \arrayrulecolor{gray!50} 
    \renewcommand{\arraystretch}{1.5} 
    \begin{tabular}{|l|c|c|c|}
        \hline
        \rowcolor{gray!25}\textbf{Key features for the experiment} & \textbf{Approach 1} & \textbf{Approach 2} & \textbf{Approach 3} \\
        \hline
        1. Used real data & $\times$ & $\times$ & $\checkmark$ \\
        \hline
        2. Generated realistic simulated data & $\times$ & $\times$ & $-$ \\
        \hline
        3. Used line fit approach & $\times$ & $\checkmark$ & $\checkmark$ \\
        \hline
        4. Both $x$ and $y$ errors considered in line fit & $\times$ & $\times$ & $\checkmark$ \\
        \hline
        5. Plotted graph with error bars & $\times$ & $\checkmark$ & $\checkmark$ \\
        \hline
        6. Correct methodology for spring constant  & $\checkmark$ & $\checkmark$ & $\checkmark$ \\
        \hline
        7. Correct error propagation & $\times$ & $\checkmark$ & $\checkmark$ \\
        \hline       
        8. Plausible spring constant (and uncertainty)& $\checkmark$ & $\checkmark$ & $-$ \\
        \hline      
    \end{tabular}
    \captionsetup{justification=raggedright, singlelinecheck=false}
    \caption{Summary of key features across three Approaches, where \checkmark and $\times$ denote ``existing" and ``non-existing", respectively. Because Approach 3 uses real student data, feature 2 and 8 were ``Not applicable". Details relating to each of these features can be found in the Discussion section.}
    \label{tab:summary}
\end{table*}

\subsection{Simulation, Analysis, and Error Propagation}

\begin{center}
\fbox{%
  \begin{minipage}{0.95\columnwidth}
    \setlength{\parindent}{10pt}
    \noindent Research Question 1: How effectively can ChatGPT Code Interpreter generate, analyse, and interpret data relating to an introductory laboratory activity?
  \end{minipage}%
}
\end{center}



A key aspect of the research project was to test whether ChatGPT Code Interpreter was able to generate realistic data for the activity and then analyse the data to determine values for the spring constant (and associated uncertainties). A summary of the key features, successes, and failures of the each Approach is provided in Table \ref{tab:summary}. 

\subsubsection{Generating realistic measurement data}

A summary of the data generation process for Approach 1 and 2 can be found in Table \ref{table:comparison2}. In both cases, simulated data was created for extensions and time periods by using $x=mg/k$ and $T=2\pi \sqrt{m/k}$ to generate `true' values of extension and time period (for an array of mass values). Noise was then added to simulate random measurement error. In Approach 1, this was achieved by setting the standard deviation of the normal distribution (from which the noise was drawn) at a fixed value of 0.02, which represented the constant relative uncertainty that had been assumed for all `measurements'. In Approach 2, the standard deviation of the normal distribution representing the noise was variable, equal to the product of the relative uncertainty ($5\%$) and the `true' value. As a result, data points with larger absolute values tended to be more dispersed around the `true' values. This heteroscedasticity in the simulated data can be clearly identified by the dispersion of the data around the line-fit for larger absolute values in both Figure \ref{fig:GraphOutput1} and Figure \ref{fig:Method2Graph}. A detailed explanation of how the data generation process in Attempt 2 led to heteroscedastic data is outlined in the Appendix.

It is worth emphasising that although both of the least squares line fits for Method 1 and Method 2 (in Approach 2) led to plausible values for the spring constant and its associated uncertainty, the data itself is not plausible due to the assumed relative uncertainties in extension and time period measurements, as well as the heteroscedastic dispersion of the data points at large absolute values. Most introductory labs require students to measure extension using a metre rule, and time periods using a stop clock. If repeat measurements are taken, then it is standard practise to estimate the `random' component of measurement uncertainty as the standard error on the mean value of the repeat measurements. In this case, it is highly unlikely that both the magnitude of the absolute uncertainty, and the dispersion of the data points around the line fit, will increase for larger absolute values. For this reason, we consider the simulated data to be `unrealistic'.

\subsubsection{Analysis, error propagation, and chi-square}

Both Approach 1 and Approach 2 resulted in spring constant values consistent with the value of $51.5 N/m$ stated in the lab activity documentation. For Method 1 of Approach 1, the spring constant was calculated directly from Hooke's Law as $k = F/x$, where $F = mg$, for each pair of force and extension measurements. The uncertainty in the spring constant was then determined using error propagation. However, as noted in the results section, the force uncertainty was incorrectly defined as a constant $2\%$ of the gravitational field strength, leading to an incorrect estimation of the uncertainty in the force. For Method 2 of Approach 1, the spring constant was correctly determined using the relation $k = 4\pi^2 m/T^2$, and the corresponding uncertainty in $k$ was correctly calculated by propagating the uncertainties $\Delta T^2$ and $\Delta m$. 

It is important to note that, although perfectly reasonable, the method suggested by ChatGPT during Approach 1 isn't typically used in introductory lab courses. Instead of using the measurements to perform a linear fit and then deriving the spring constant and its uncertainty from the line fit parameters (gradient and uncertainty in the gradient), ChatGPT calculated the spring constant and its uncertainty for each measurement pair independently. The approach suggested by ChatGPT could have been improved if the mean and standard error on the mean of the individual spring constants had then been calculated, but this step was not taken.

In Approach 2, ChatGPT was explicitly prompted to \textit{``plot a graph with error bars''} and \textit{``perform a chi-squared fit''}, with ChatGPT opting for a least squares line fit using Python's \texttt{curve\_fit} function. However, this line fit approach only included $y$-errors (extension uncertainties and $T^2$ uncertainties), and assumed that the uncertainties in the independent variables (Force and mass) were negligible by comparison. However, if the $x$-errors are non-negligible, as was the case in both Method 1 and Method 2 (as can be seen by the size of the error bars in Figure \ref{fig:GraphOutput1} and Figure \ref{fig:Method2Graph}), then a more appropriate approach would have been for ChatGPT to use an Orthogonal Distance Regression (ODR) line-fit that takes into account both $x$-errors and $y$-errors. However, despite the limitations of the line-fitting approach, the gradient and uncertainty in the gradient resulting from the \texttt{curve\_fit} line-fit were correctly used to determine values for the spring constant and uncertainty in the spring constant.

In Approach 3, real student data was uploaded and ChatGPT was prompted to \textit{``perform an ODR fit, then use the fit data to calculate the spring constant and uncertainty in spring constant''}. For both Method 1 and Method 2, the ODR fit parameters were used to correctly calculate the spring constant and its uncertainty, and these values matched the values obtained by the researchers when performing the analysis on the real student data. In both Approach 2 and Approach 3, ChatGPT was able to correctly plot the data (including error bars), with the fit line overlaid on the plotted data points. 

Finally, the reduced chi-square statistic was correctly calculated using the model fit parameters and uncertainties in both Method 1 and Method 2 (although it is again worth mentioning that in Approach 2, only the $y$-errors were included in the fit and calculation of the reduced chi-square statistic). However, in Approach 3, having provided the initial prompt to use an ODR fit, the reduced chi-square statistic took into account uncertainties in both the $x$ and $y$ variables.

\subsection{Providing Prompts to ChatGPT}

\begin{center}
\fbox{%
  \begin{minipage}{0.95\columnwidth}
    \setlength{\parindent}{10pt}
    \noindent Research Question 2: What strategies can be implemented to optimise effective communication with ChatGPT during laboratory tasks to ensure high quality, reliable output?
  \end{minipage}%
}
\end{center}

The evolution of our interactions with ChatGPT across Approach 1, 2 and 3 revealed that the specificity and structure of the prompts plays a crucial role in guiding ChatGPT's responses, as does the addition of context and detail. We also found that by engaging in follow-up questions and asking for clarification, the quality and relevance of ChatGPT's output improves. In summary:

\begin{itemize}
\item \textbf{Specificity Matters:} A detailed and specific initial prompt will often lead to a more targeted and comprehensive reply. By providing as much relevant information as possible in the initial prompt, fewer follow-on questions and clarifications will be needed. For example, the initial prompts provided during Approach 2 led to a more comprehensive analysis than Approach 1..

\item \textbf{Iterative Questioning:} By gradually refining the prompts and iterating on previous interactions, the quality and relevance of response is improved. By actively querying and questioning ChatGPT responses, mistakes in output can be more easily identified and corrected. 

\item \textbf{Explicit Instruction:} By clearly stating the desired format or approach to a given task, ChatGPT is more likely to present you with relevant and useful information. For example, explicitly stating the required format for the column headings in a table, or which variables should be plotted on the $x$ and $y$ axes, or how the errors should be propagated through an equation. 

\item \textbf{Avoid Ambiguity and Assumptions:} In the absence of clear and specific instructions, ChatGPT is likely to make assumptions that are not explicitly stated or justified. For example, in Approach 1, without specifying \textit{how} we wanted the data to be generated and analysed, ChatGPT proceeded to calculate the spring constant using pairs of values, rather than a graphical line-fitting approach.

\end{itemize}

\subsection{Implications for Laboratory Classes}

\begin{center}
\fbox{%
  \begin{minipage}{0.95\columnwidth}
    \setlength{\parindent}{10pt}
    \noindent Research Question 3: What are the implications of AI tools like ChatGPT Code Interpreter on introductory laboratory courses?
  \end{minipage}%
}
\end{center}

\textbf{Student Implications:} 
ChatGPT has the potential to help bridge the gap between what a student can accomplish independently, and what they can achieve with targeted expert support \cite{vygotsky1962development}, especially in beginning laboratory settings where ChatGPT can support data modelling, analysis, and interpretation. However, our experience in this study suggests that novice learners might struggle to identify ChatGPT errors, or to provide the necessary level of prompt detail when questioning ChatGPT. High quality ChatGPT output requires highly detailed and expert-like prompts, as well as a critically trained eye that is capable of spotting potential errors and scrutinising Python code to better understand the cause of the underlying problem. Even with Code Interpreter's remarkable powers, users must still exercise caution and critical thinking when interpreting data analysis outputs. However, despite the challenges, there is still scope to design tailored ChatGPT activities that encourage students to think critically about the questions they pose, iterate on previous interactions, and scrutinise the answers provided by ChatGPT. These are precisely the types of scientific skills that we should be encouraging and developing in our students. 

Given that many students are already using ChatGPT without expert guidance and support \cite{Intelligent2023}, it is essential to be proactive and provide students with training sessions and guidelines on how to use ChatGPT effectively and ethically as a tutoring resource. It is the responsibility of educational institutions to ensure that students are well-informed about how to use AI tools in a responsible manner.

\textbf{Instructor Implications:} ChatGPT Code Interpreter is an exciting new tool that provides a variety of opportunities for university instructors. In addition to the important role of ensuring equitable and ethical use of ChatGPT, instructors can also use ChatGPT to help prepare unique, challenging, and tailored problem-solving activities that help to engage students. Furthermore, Code Interpreter can be used to generate simulated data for lab activities, create plots and visualisations of existing data, and perform simple calculations. Code Interpreter can be used to design interactive slider-based simulations and animations that can help students to visualise and conceptualise challenging topics (such as the relationship between error bars, line fits, and the reduced chi-square statistic). 

Code Interpreter also has the potential to save time and effort in preparing teaching materials and provide instructors with the opportunity to provide quick, visual representations of data. Code Interpreter also allows for the export of data files, images, and python scripts, allowing for the quick distribution and sharing of information. However, instructors should be aware of the limitations and potential errors that may arise when using ChatGPT, as highlighted in this article. It is essential that instructors scrutinise the output generated by ChatGPT to ensure accuracy and reliability, especially if generating resources for use by students.

Although it is not discussed in this paper, the authors have initiated a new research study in which ChatGPT Code Interpreter is used to generate novel and exciting lab activities. Initial results are encouraging in terms model data generation, detailed lesson plans, marking rubrics, and synoptic extension activities. We hope to report further on this in the near future.

\section{Limitations of ChatGPT and Future Directions}

Understanding the limitations of AI tools is essential when considering their use and integration into introductory labs. While ChatGPT was able to generate plausible-looking values for the spring constant (and associated uncertainty) in Approach 1,2 and 3, there were several limitations and challenges observed. As already noted, the random-noise-adding process in Attempt 2 led to heteroscedastic data. This issue could be addressed by providing a greater level of specificity in the prompts relating to how the measurement errors should be approximated. However, it is unlikely that a novice learner would spot such an error or know how to address it.

Another issue worth mentioning is that Code Interpreter will sometimes produce significantly different outputs when provided with the same identical prompt input. For example, during our initial investigations we uploaded the same lab activity via Code Interpreter into four separate ChatGPT sessions, and provided an identical initial prompt relating to data generation and graph plotting. Three of the four plots looked reasonable, giving rise to a spring constant and uncertainty consistent with the stated value 51.5 N/m. However, one of the plots gave rise to a spring constant of 36859.69 N/m. This observation highlights that while Code Interpreter can often reproduce similar data and results when provided with identical prompts, there are instances where this is certainly not the case. This is yet another example that highlights the importance of sense-checking and scrutinising the output provided by ChatGPT.

Given students' varying levels of expertise and familiarity with lab work, ChatGPT, and Python, there is a risk that students will not be able to identify errors in Code Interpreter's output, or be able to effectively use Code Interpreter to tackle specific physics problems. This could lead to misconceptions in their understanding of key concepts. Future studies should therefore focus on examining how students interact with ChatGPT Code Interpreter to solve physics problems. 

Despite the relative success of Code Interpreter in generating and analysing data for this introductory lab activity, its effectiveness in handling more advanced or inquiry-based laboratory activities remains uncertain. However, considering the rapid speed of AI development, we expect that that future iterations of ChatGPT will be increasingly capable of addressing more complex experimental scenarios in university-level physics education.

\section{Appendix: Heteroscedasticity in simulated data} \label{subsec:heterosc}

An interesting feature of the data generation process in Appproach 2 was the fact that ChatGPT opted to simulate random measurement error by:
\begin{itemize}
    \item Assuming a relative uncertainty in all measurements of mass, force, extension, and time period.
    \item Adding random noise to each measurement with the noise drawn from a normal distribution with a standard deviation equal to the product of the relative uncertainty and the measurement value.
\end{itemize}

The result of these two assumptions is that:

\begin{figure*}[tp]
    \centering
    \includegraphics[scale=0.52]{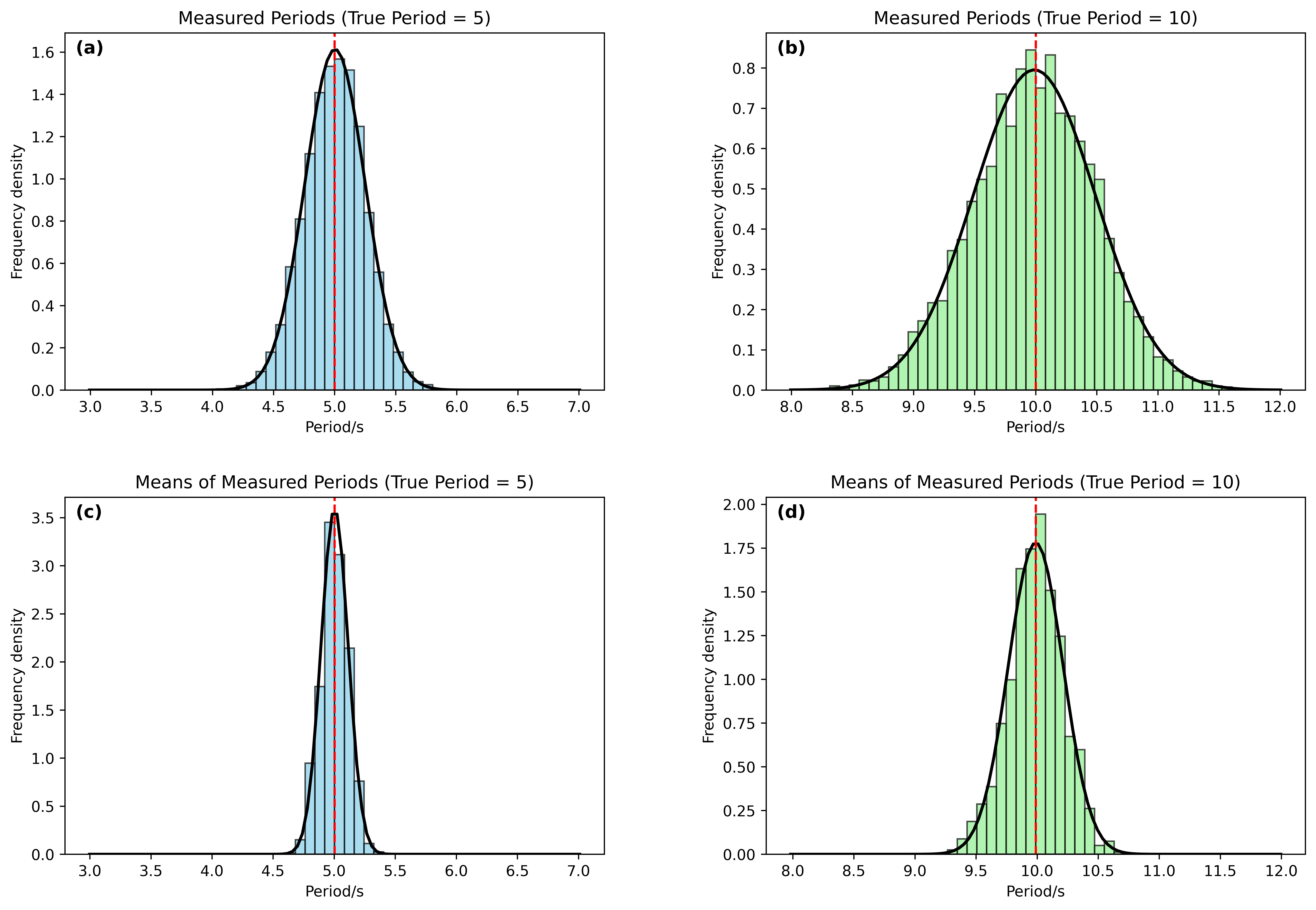}
    \captionsetup{justification=raggedright, singlelinecheck=false}
    \caption{The top histograms (fig. a and b) show the distributions of period values for a true period (indicated by the red dashed line) of 5s and a true period of 10s, assuming that the standard deviation of the distribution is equal to $5\%$ of the true value. The bottom histograms (fig. c and d) show the mean values of the measured periods.}
    \label{fig:Historgrams}
\end{figure*}

\begin{itemize}
    \item The size of the error bar for each measurements is proportional to the magnitude of the measurement value. 
    \item The data is heteroscedastic. This means the dispersion of the plotted points around the line-fit is greater for larger measurement values. In other words, the variance of the errors increases with the value of the independent variable. 
\end{itemize}

We can see this explicitly by looking at how the data was simulated for the force-extension graph in Figure \ref{fig:GraphOutput1} during Approach 2. As mentioned earlier, the `true' values of extension were calculated from the simulated mass values using Hooke's Law:

\[ x = \frac{mg}{k} \]

Next, following ChatGPT's approach, we add a noise variable, $\epsilon$, to the extension measurements. The noise variable $\epsilon$ is chosen to follow a standard normal distribution, meaning it has a mean of zero and a standard deviation of one. We can scale the standard deviation to be equal to the product of the the relative uncertainty, $u$, of the extension measurements and the true extension value, $x$, by multiplying $\epsilon$ by $ux$. It then follows that the simulated `measured' extension, $x_{\text{measured}}$, will be equal to:

\[ x_{\text{measured}} = x+x\epsilon u = x(1 + \epsilon u) \]

Next, imagine repeating this data generation process a very large number of times, and then calculating the variance of the simulated extension measurements:

\begin{align*}
\text{Var}(x_{\text{measured}}) &= \text{Var}\left(x\left(1 + \epsilon u\right)\right) \\
&= x^2 u^2
\end{align*}
where we made use of the fact that the `true' extension, $x$, is a constant, and that $\text{Var}(cX) = c^2 \text{Var}(X)$ where $c$ is a constant, $\text{Var}(X + Y) = \text{Var}(X) + \text{Var}(Y)$ when $X$ and $Y$ are independent, and $\text{Var}(\epsilon) = 1$. It then follows that the standard deviation of the simulated measurement values would be equal to the square root of the variance:
\begin{align*}
\sigma(x_{\text{measured}}) &= \sqrt{x^2 u^2} = xu 
\end{align*}

So we see that the standard deviation of the simulated measured extension is proportional to the true extension, with larger values of measured extension more dispersed around the true value than smaller values. In other words, as the extension (and mass) increases, the variability of the measured extensions also increases, resulting in a heteroscedastic pattern of the data as observed in Figure \ref{fig:GraphOutput1}.

We can follow a similar procedure to show that the data generation process suggested by ChatGPT leads to heteroscedastic data in  $T$ (and therefore $T^2$), as clearly evident in Figure \ref{fig:Method2Graph}. If we recall, ChatGPT began by generating equally spaced mass values, then used $T = 2\pi\sqrt{\frac{m}{k}}$ to generate `true' period values. Random noise was then added to these true period values to generate the `measured' period:

\[ T_{\text{measured}} = T+T\epsilon u = T(1 + \epsilon u) \]

Following an identical argument to that used when considering $x_{\text{measured}}$, it is simple to show that:

\begin{align*}
\sigma(T_{\text{measured}}) &= \sqrt{T^2 u^2} = Tu 
\end{align*}

For each true period, $T$, ChatGPT generated five values of $T_{\text{measured}}$ and then calculated a mean period $T_{\text{mean}}$. Let's consider the standard deviation of $T_{mean}$. The first step is to write:

\begin{align*}
T_{\text{mean}} = \frac{1}{5} \sum_{i=1}^{5} T(1 + \epsilon_i u)
\end{align*}

It then follows that:
\begin{align*}
\text{Var}(T_{\text{mean}}) &= \text{Var}\left(\frac{1}{5} \sum_{i=1}^{5} T(1 + \epsilon_i u)\right) \\
&=\frac{T^2}{25} \sum_{i=1}^{5} \text{Var}\left((1 + \epsilon_i u)\right) \\
&= \frac{T^2 u^2}{5}
\end{align*}

The standard deviation of the mean time period is then:
\begin{align*}
\sigma(T_{\text{mean}}) = \sqrt{\text{Var}(T_{\text{mean}})} = \frac{Tu}{\sqrt{5}}
\end{align*}

This result confirms that the standard deviation of the mean of five repeat measurements is proportional to the `true' time period, and hence for larger time period values, we expect the plotted data to be more dispersed. We also see that the standard deviation of the mean of the 5 measured time periods is reduced by a factor of $\sqrt{5}$ compared to the standard deviation of a single measured time period. This reduction is due to the averaging process, which tends to cancel out random fluctuations, resulting in a more precise estimate of the true value. What we have shown is a special case of the common expression for the standard deviation of the mean (standard error of the mean):
\begin{align*}
\sigma(\bar{X}) = \frac{\sigma(X)}{\sqrt{n}}
\end{align*}
where $\sigma(X)$ is the standard deviation of a single measurement, $\sigma(\bar{X})$ is the standard deviation of the mean, and $n$ is the number of measurements. A visualisation of this effect is shown in the histograms appearing in Figure \ref{fig:Historgrams}. We see that the width of the distribution for the measured period increases as we move from $T=5s$ to $T=10s$. Furthermore, we see that for both time periods, the width of the distribution representing the mean time period is smaller than the width of the distribution representing a single measurement. Furthermore, we see that the width is reduced by a factor of $\sqrt{5}$. Finally, we note that mean distribution is wider when the true time period is 10s compared to 5s, again demonstrating that the data generated via this approach will be heteroscedastic.


\bibliography{ms.bib}

\end{document}